\documentclass[12pt,makeidx,epsfig,wrapfig,amsbsy,amsfonts]{article}
\usepackage[T2A]{fontenc}
\usepackage{times}
\usepackage{amsmath}
\usepackage{graphicx}
\usepackage{epsfig}
\advance \topmargin by -\headheight
\advance \topmargin by -\headsep
\evensidemargin \oddsidemargin
\begin{document}
\pagestyle{plain}

\begin{titlepage}
\vspace*{5.15cm}
\begin{center}
{\Large\bf
    Measurement of the $K^{-} \rightarrow \pi^{0} e^{-} \bar \nu(\gamma) $ branching
     ratio }
\vspace*{0.5cm}

{\bf  V.I.~Romanovsky, S.A.~Akimenko, G.I.~Britvich, K.V.~Datsko,  A.P.~Filin, 
A.V.~Inyakin,  A.S.~Konstantinov, I.Y.~Korolkov, V.A.~Khmelnikov,
V.M.~Leontiev, V.P.~Novikov, V.F.~Obraztsov,  V.A.~Polyakov,
 V.I.~Shelikhov,  O.G.~Tchikilev, V.A.~Uvarov,  O.P.~Yushchenko. }
\vskip 0.15cm
{\bf $Institute~for~High~Energy~Physics,~Protvino,~Russia$}
\vskip 0.35cm
{\bf V.N.~Bolotov, V.A.~Duk, S.V.Laptev,  A.Yu.~Polyarush. }
\vskip 0.15cm
{\bf $Institute~for~Nuclear~Research,~Moscow,~Russia$}
\end{center}
\end{titlepage}
\newpage

\vskip 5.0cm
\begin{center}
Abstract
\end{center}

The  branching fraction for the decay $K^{-} \rightarrow \pi^{0} e \bar \nu$ 
is  measured using in-flight decays detected with {\bf ISTRA+} setup working 
at the 25 GeV negative secondary beam of the U-70~PS:\linebreak 
$Br_{K_{e3}}= (5.124 \pm 0.009 (stat) \pm 0.029(norm) \pm 0.030(syst))\%$. 

From this value   the $|V_{us}|$ element of the CKM matrix is extracted,
using previously measured $f_{+}(t)$ form factor:
$|V_{us}|=0.227 \pm 0.002$. The results are in agreement with recent 
measurements by BNL E865, FNAL KTeV, KLOE .

\newpage
\sloppy

\section{ Introduction}
 
The decay $K \rightarrow e \nu \pi^{0}$($K_{e3}$) is known to be one of the
best sources of information about $V_{us}$ element of the
Cabibbo-Kobayashi-Maskawa (CKM) matrix. The interest in high statistics, low
systematics measurement of the $K_{e3}$ branching has raised after the paper
by BNL E865 collaboration \cite{Poblaguev}, where 2.5~$\sigma$ increase of 
the $K^{+}_{e3}$ branching as compared with PDG02 \cite{PDG02} was reported. 
This result improved the agreement of the measured $V_{ud},
V_{us}, V_{ub}$ with the unitarity condition:
\begin{displaymath}
|V_{ud}|^{2}+|V_{us}|^{2}+|V_{ub}|^{2}=1\ ,
\end{displaymath}
which was violated by 2.3 $\sigma$ with the old value of $V_{us}$.
Since then, a set of new measurements of the $K_{e3}$ branchings for $K_L$
\cite{KTeV,NA48,KLOE1}, $K_S$\cite{KLOE2} 
has appeared, confirming the increase of $V_{us}$   value.    
In our analysis,  
we present a new measurement of $K^{-}_{e3}$ branching based on statistics    
of about 2M  events using new approach, which allows to significantly reduce the systematics uncertainties.

\section{ Experimental setup}
The experiment has been performed at the IHEP 70 GeV proton synchrotron U-70.
The experimental setup ISTRA+ (Fig.~1) has been described in some details
elsewhere~\cite{ISTRA}. 
\begin{center}
\begin{figure}[h]
\includegraphics[scale=.6 , angle=90]{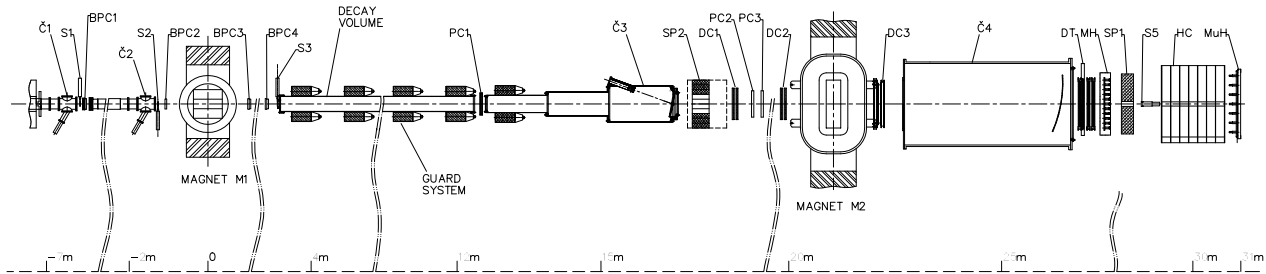}
\caption{ Elevation view of the ISTRA+ detector.}
\end{figure}
\end{center}
 The setup is located in the  negative unseparated secondary beam. 
The beam momentum in the measurements is $\sim 25$ GeV with 
$\Delta p/p \sim 1.5 \% $. The admixture of $K^{-}$ in the beam is $\sim 3 \%$.
The beam intensity is $\sim 3 \cdot 10^{6}$ per 1.9 sec. U-70 spill.
The  beam particle deflected by M$_{1}$ is 
measured by $BPC_{1}\div BPC_{4}$ PC's
with 1mm wire step, the kaon identification is done by 
$\check{C_{0}} \div \check{C_{2}}$ threshold $\check{C}$-counters. 
The 9 meter long vacuumed
decay volume is surrounded by 8 lead glass rings $LG_{1} \div LG_{8}$ used to
veto low energy photons. $SP_{2}$ is a lead glass calorimeter to detect/veto 
large angle photons.
The charged decay products deflected in M2 with 1~Tm field integral
are measured with the help of $PC_{1} \div PC_{3}$ --- 2~mm step proportional chambers;
$DC_{1} \div DC_{3}$ --- 1~cm cell drift chambers and finally with 2~cm diameter
drift  tubes    $DT_{1} \div DT_{4}$. 
Wide aperture threshold Cerenkov counters $\check{C_{3}}$, 
$\check{C_{4}}$ are filled  with He and
are not used in the present measurements. $SP_{1}$ is a 576-cell lead glass calorimeter,
followed by HC --- a scintillator-iron sampling hadron calorimeter, subdivided
into 7 longitudinal sections 7$\times$7 cells each. MH is a 
11$\times$11 cell scintillating hodoscope, used to  improve the time resolution of the 
tracking system, MuH  is a  7$\times$7 cell muon hodoscope. 

 The trigger is provided by $S_{1} \div S_{5}$ scintillation counters, 
$\check{C_{0}} \div \check{C_{2}}$ Cerenkov counters,
analog sum of amplitudes from the last dinodes of the $SP_1$ :
 $T=S_{1} \cdot S_{2} \cdot S_{3} \cdot 
 \bar{S_{4}} \cdot \check{C_{0}} \cdot \bar{\check{C_{1}}} \cdot 
 \bar{\check{C_{2}}} \cdot 
 \bar{S_{5}} \cdot \Sigma(SP_{1})$,
here $S_4$ is a scintillator counter with a hole to suppress beam halo;
 $S_5$ is a counter downstream  the setup at the beam focus;
$\Sigma(SP_{1})$ --- a requirement for the analog sum of amplitudes from 
$SP_1$ to be larger than $\sim$700 MeV --- a MIP signal. The last requirement 
serves to suppress the $K \rightarrow \mu \nu$ decay. 

\section{General description of  the experimental method}
Our experimental approach to the $K_{e3}$ branching ratio  measurement is based on the 
following points:
\begin{enumerate}
\item $K_{e3}$ is the dominant source of  electrons in single track decays of charged kaon.
Indeed, $Br_{K_{e3}} \sim 5 \% ; Br_{K_{e2}} \sim 1.5 \times 10^{-5}; 
Br_{K_{e2 \gamma}}\sim 1.5 \times 10^{-5} ; Br_{K_{e \nu \pi^{0} \pi^{0}}} 
\sim 2 \times 10^{-5}$. The  contribution from the decay chain $K \rightarrow \mu \nu
;\ \mu \rightarrow e \nu \bar \nu$ corresponds to effective  $Br_{K_{\mu2}} < 10^{-5}$, because
of the long lifetime of muon. The background sources do not exceed fraction of $\%$ from
$K_{e3}$ and can be easily taken into account. 
\item The number of electrons is obtained from the 
fit of the $E/p$ distribution, where $E$ is the energy of the shower, associated with the 
charged track with momentum $p$.
\item The decay $K^{-} \rightarrow \pi^{-} \pi^{0}(K_{\pi2})$, which is used for the  
normalization is identified by the peak in the distribution over momentum of the 
charged secondary track in the kaon c.m.s ($p^{cms}_{\pi}$---distribution). The number
of  $K_{\pi2}$ events is obtained from the fit of this distribution.
\item This method is based on the reconstruction of the beam and decay track only, i.e 
does not require a reconstruction of $\pi^0$ both in $K_{e3}$ and
$K_{\pi 2}$ decays. It uses few selection cuts, thus one can hope for a small systematics
in this analysis. 
\end{enumerate}

\section{Data set and event selection}

During physics run in Winter 2001,
332M events were logged on tapes. This information is complemented
by about 260M MC events generated with Geant3 \cite{geant} Monte Carlo program.
MC generation includes a realistic description of the setup including decay volume
entrance windows, track chambers windows, gas, sense wires and cathode structure,
Cerenkov counters mirrors and gas, shower generation in EM calorimeters, etc.

 The data processing starts with the beam particle reconstruction in 
$BPC_{1} \div BPC_{4}$, then the secondary tracks are looked for in 
$PC_{1} \div PC_{3}$ ; $DC_{1} \div DC_{3}$; $DT_{1} \div DT_{4}$
and events with one good negative track are selected.
The decay vertex is searched for, and a cut $P>1 \%$ on the probability of 
the vertex fit is introduced. 

The decay vertex is selected to locate in the
decay volume region $500 <z <1500$ cm, $z$ being the coordinate along the beam line.
To suppress undecayed particles(beam electrons, in particular) a cut on the space
angle between beam and secondary track is introduced: $\Delta \theta> 2$
mrad. 

The next step is to require the total energy in the 
$SP_{1}$ calorimeter to be above one GeV: $E_{SP1} > 1$ GeV. 
This cut repeats 
``digitally'' the trigger
requirement, which is introduced to suppress $K\rightarrow \mu \nu $ decays. 
The matching of the charged track and a shower in $SP_{1}$ is done 
on the basis of the distance {\bf r} between the track extrapolation to the 
calorimeter front surface and the shower coordinates ($r\leq 5$ cm). This cut is used for
 the electron identification only. 
 
\section{Verification  of the method on Monte-Carlo events.} 
The cuts described above
were applied to the MC-sample which contains a natural 
mixture of reconstructed six largest kaon decays ($\mu^- \nu$, $\pi^- \pi^0$, $\pi^- \pi^+ \pi^-$, $\pi^0 e^- \nu$, $\pi^0 \mu^- \nu$, $\pi^-\pi^0\pi^0 $), i.e the sample includes both
signal and main backgrounds. \\ [-2mm]

\begin{minipage}[t]{8.cm}
\includegraphics[width=8.0 cm]{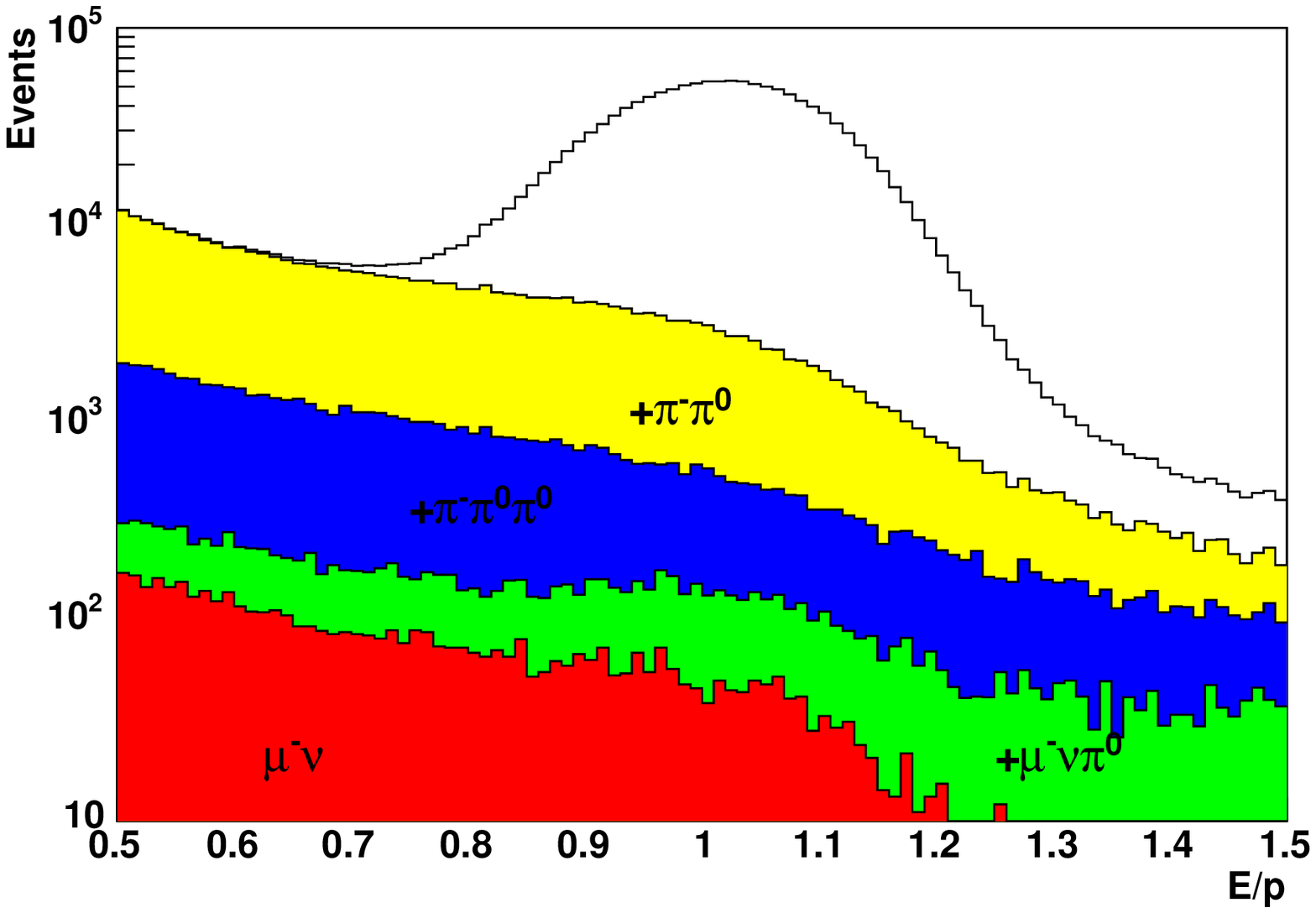} 
\begin{center}
{\small Figure 2: The cumulative distributions over the ratio of the energy of the associated calorimeter
 cluster to the momentum of the charged track $(E/p$ plot) for four largest background decays
 and $K_{e3}$ signal events (MC-events).}
\end{center}
\end{minipage} \ \hfill \ 
\begin{minipage}[t]{8.cm}
\includegraphics[width=8.0 cm]{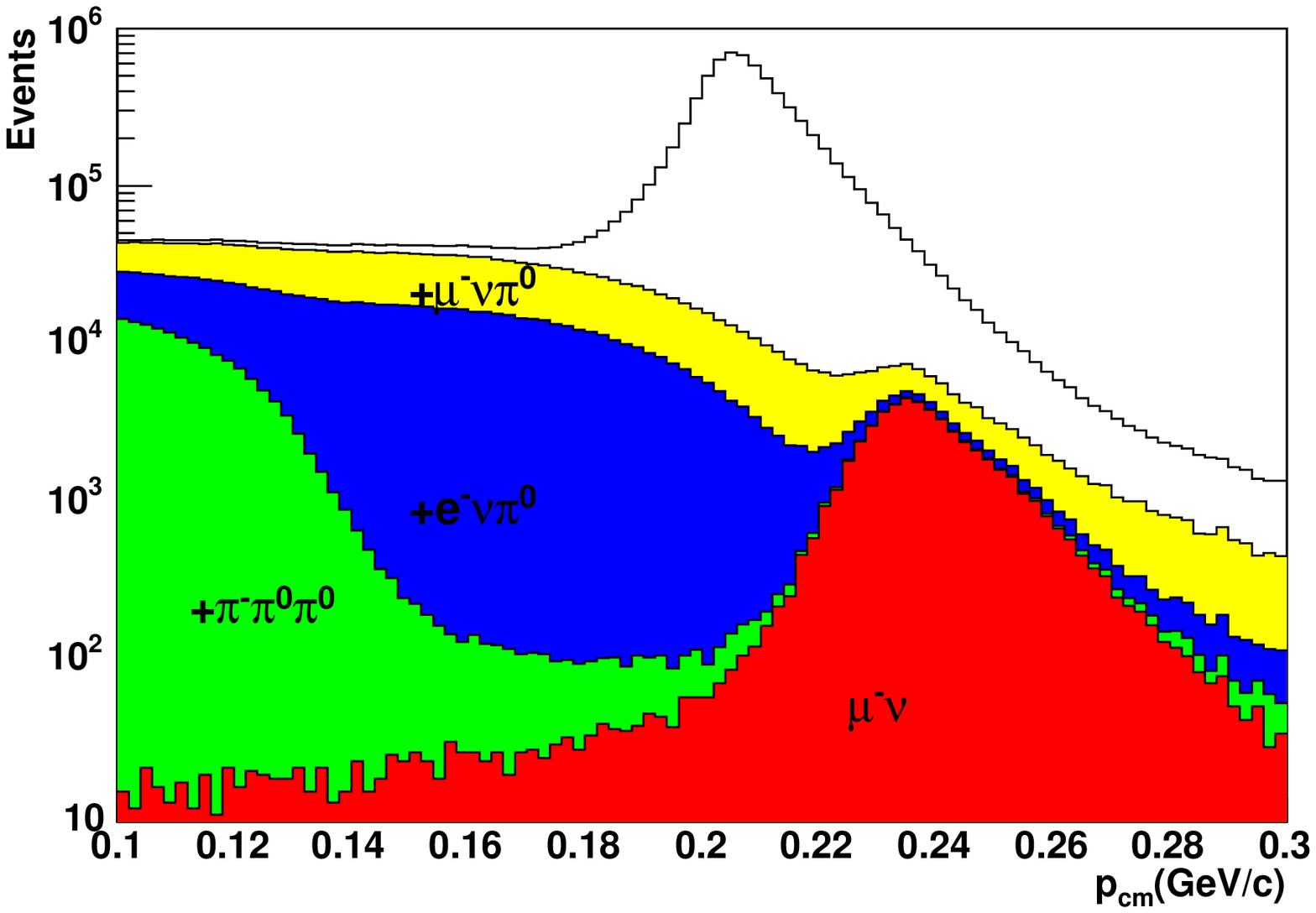} 
\begin{center}
{\small Figure 3: The cumulative distributions over $p^{cms}_{\pi}$~--- the momentum of
the secondary particle in the kaon c.m.s. system, assuming that the particle is $\pi$-meson  for four largest background decays and $K_{e3}$ 
signal events (MC-events).}
\end{center}
\end{minipage}
\vspace*{0.5cm}
\linebreak 
The $E/p$ distribution for these events is presented in Fig.~2. 
MC shows that the main background to $E/p$ is from $\pi^{-} \pi^{0}$ 
and $\pi^{-} \pi^{0} \pi^{0}$. Background is smooth enough to be described by 
$A \times e^{-P1 \cdot x}$, signal is described by sum of two 
Gaussians. Direct test of the fit
gives  $N_{K_{e3}}^{fit}=1.006 \times N_{K_{e3}}^{true}$, where $N_{K_{e3}}^{fit}$
--- is the
number of events in the peak of E/p distribution of Fig.~2 and  $N_{K_{e3}}^{true}$ is
the ``true'' number of $K_{e3}$ events in Fig.~2 known from MC.

The $p^{cms}_{\pi}$ distribution for the same events is presented in Fig.~3. 
For the $p^{cms}_{\pi}$ the main background is $e \nu \pi^{0}$
and $\mu \nu \pi^{0}$. Background is smooth enough to be described by 
4-th order polynomial. Signal is described by the sum of two Gaussians. 

\section{Data Analysis}

\begin{minipage}[h]{8.cm}
\includegraphics[width=8.0 cm]{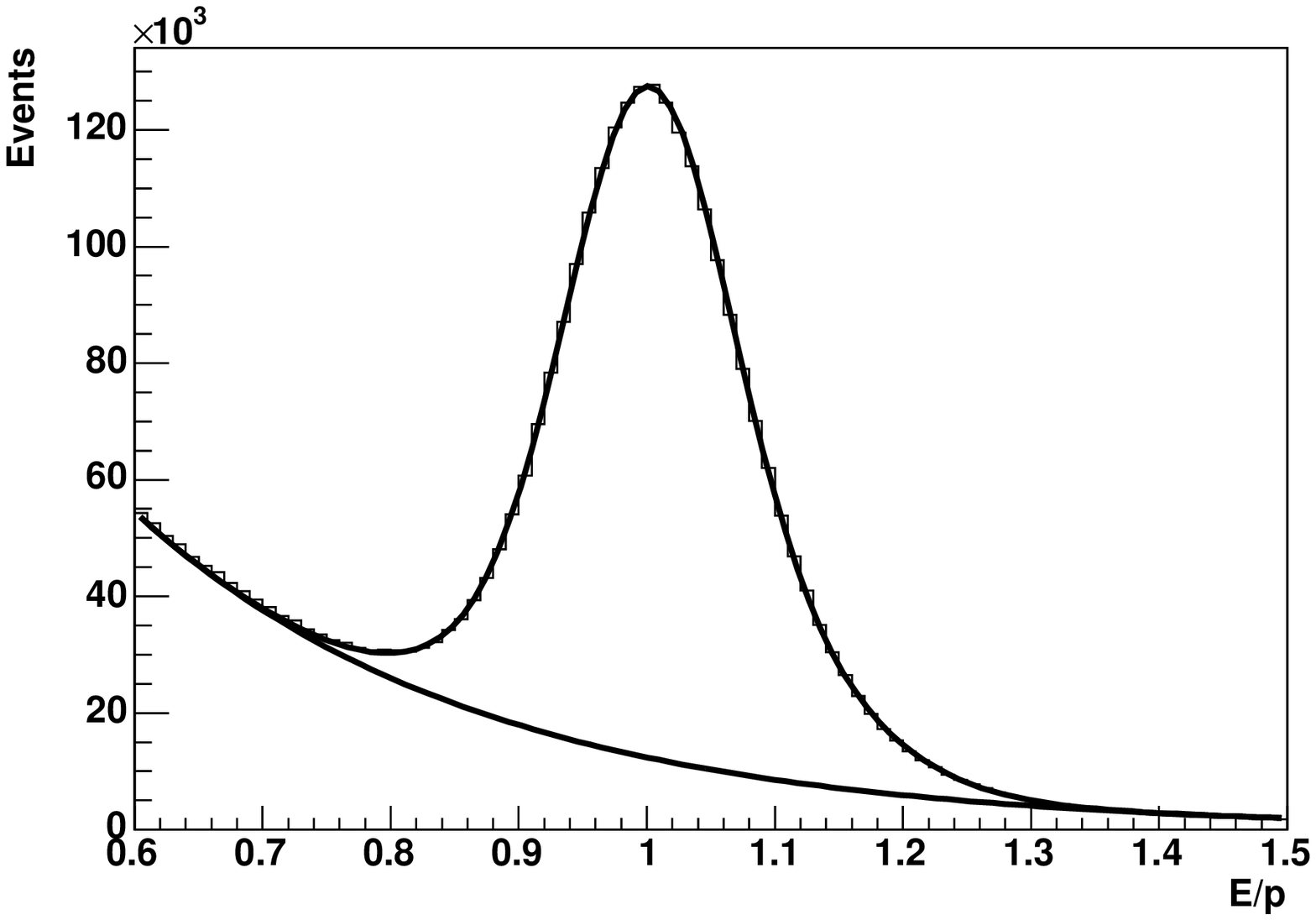} 
\begin{center}
{\small Figure 4: The  E/p distribution  for the real data.}
\end{center}
\end{minipage} \ \hfill \ 
\begin{minipage}[h]{8.cm}
\includegraphics[width=8. cm]{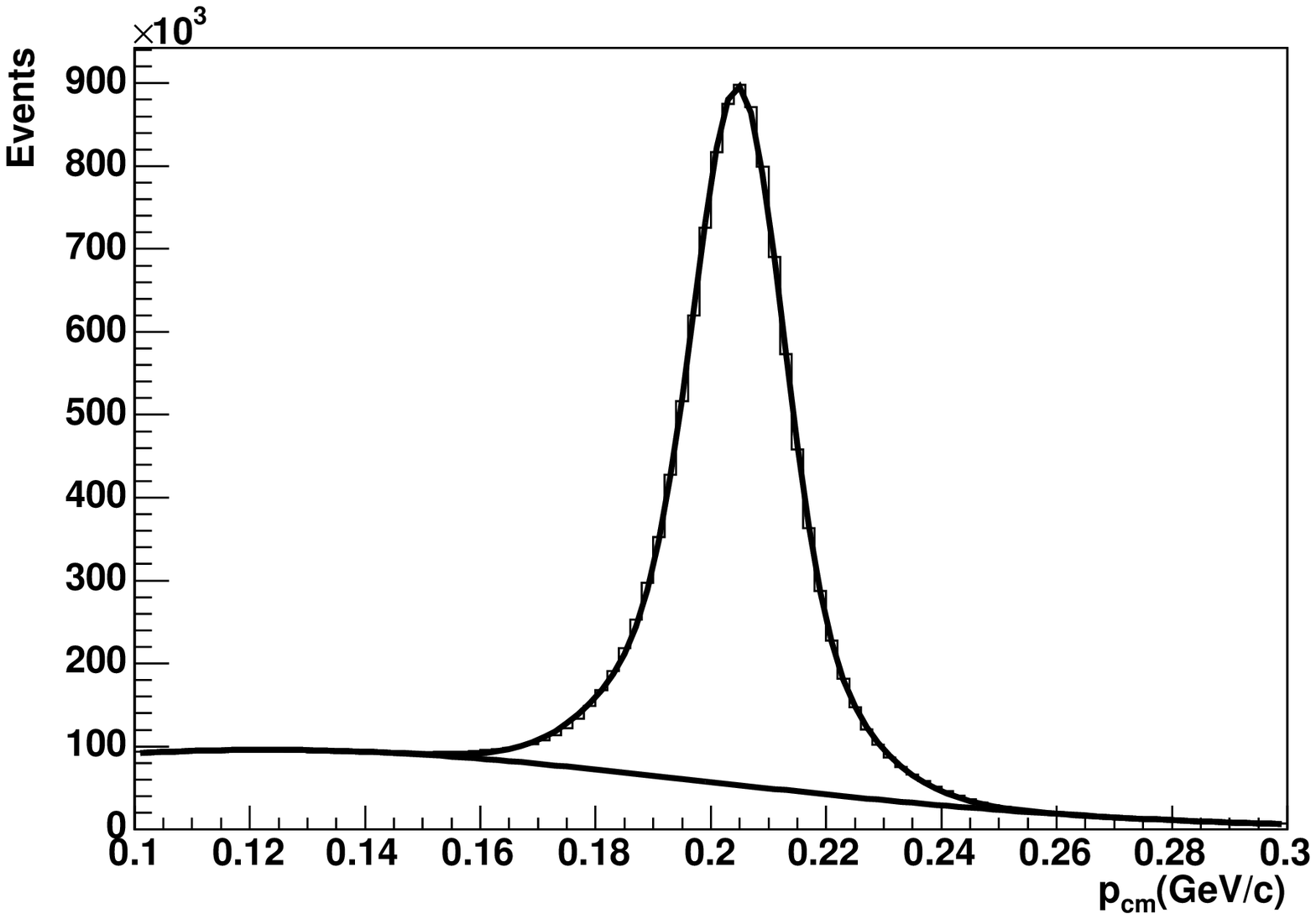} 
\begin{center}
{\small Figure 5: The distributions over the $p^{cms}_{\pi}$
 for the real data.}
\end{center}
\end{minipage}\\
\vspace*{0.5cm}
\linebreak 

The application of the procedure described above to the real data  results in 
the $E/p$ and $p^{cms}_{\pi}$ distributions of Fig.~4, Fig.~5. 
The fit of the distributions  gives: 
\begin{itemize}
\item[]
$N_{K_{e3}}=(2.1739 \pm 0.0024) \times 10^{6}$; \
$N_{K_{\pi2}}=(10.2940 \pm 0.0053) \times 10^{6}$ for the data 
\item[]
$N_{K_{e3}}=(1.2319 \pm 0.0013) \times 10^{6}$; \
$N_{K_{\pi2}}=(6.2758 \pm 0.0030) \times 10^{6}$ for the MC. 
\end{itemize}
In Geant3 version which we are using, the following branchings are assumed: 
$Br_{K_{e3}}=4.82\% ;\  Br_{K_{\pi2}}=21.17 \% $.
From this we can get:
\begin{displaymath}
 Br_{Ke3}/Br_{K_{\pi2}}=0.2449 \pm 0.0004(stat).
\end{displaymath}
Using PDG06 \cite{PDG06} value $Br_{K{\pi2}}=(20.92 \pm 0.12) \%$:
\begin{displaymath}
Br_{Ke3}=(5.124 \pm 0.009(stat) \pm 0.029(norm))\%. 
\end{displaymath}
In fact, the $K_{\pi2}$ branching of \cite{PDG06} is obtained by the fit, which has many inputs, including $Br_{K_{e3}}/Br_{K_{\pi2}}$ ratio.
That is, it would be more correct to repeat the fit with our new result on branching ratio. This is done in \cite{Moulson} together  with
averaging over all recent experimental data. In present paper we however decided to limit ourself to our own results. 

\section{Study of systematics}
The specific feature of our measurements is that the statistical error is much
smaller than the systematic one. This allows us to study systematics by
subdividing our statistics in parts over different variables. Fig.~6 shows the
dependence of the measured $N_{K_{e3}}/N_{K_{\pi2}}$ ratio versus run number and 
Fig.~7 versus $z$ --- the vertex coordinate along the beam line.\\
\begin{minipage}[h]{8.cm}
\includegraphics[width=7.5 cm]{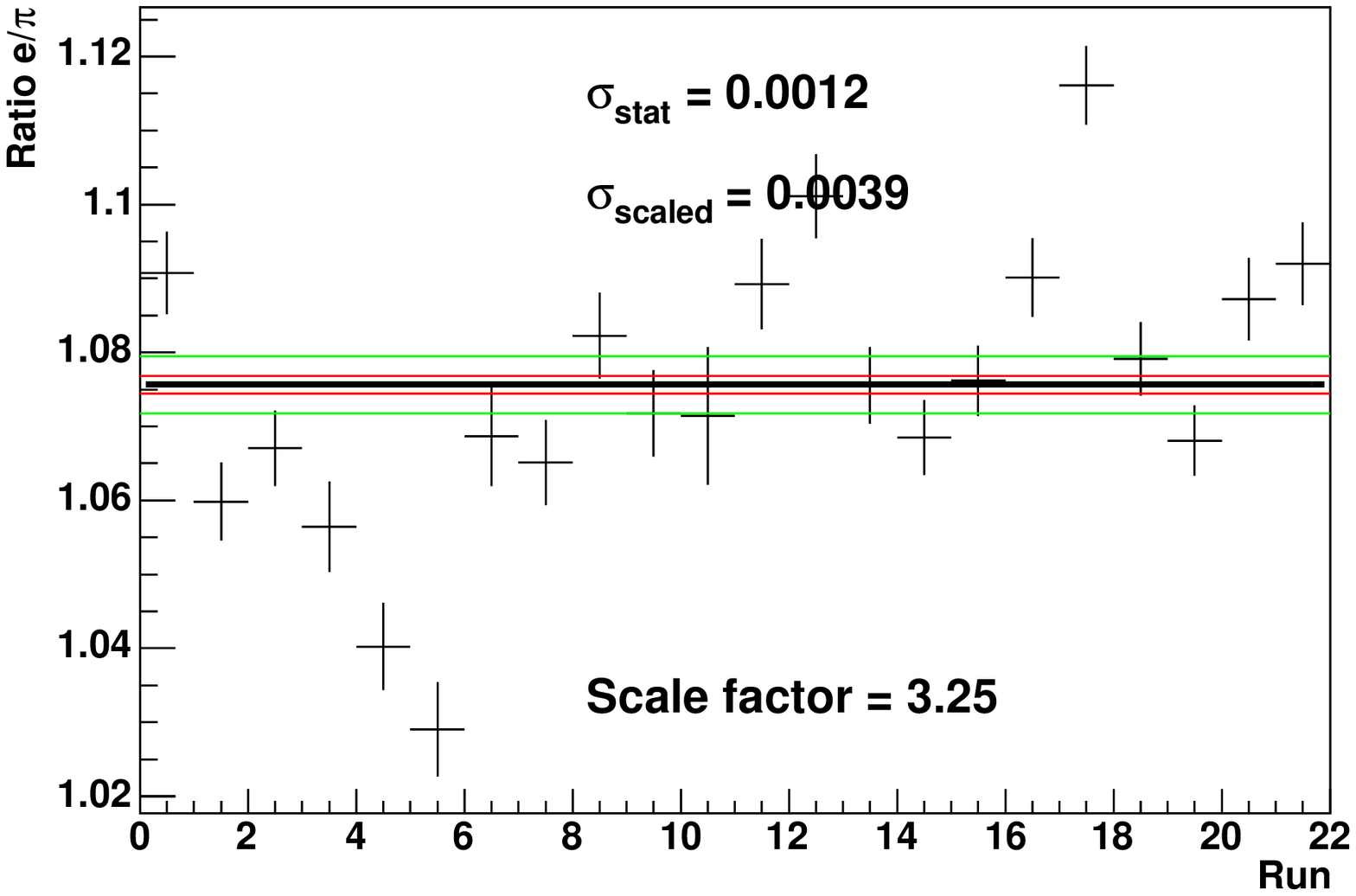} 
\begin{center}
{\small Figure 6: The  measured ratio $N_{K_{e3}}/N_{K_{\pi2}}$ versus run number.}
\end{center}
\end{minipage} \ \hfill \ 
\begin{minipage}[h]{8.cm}
\includegraphics[width=7.5 cm]{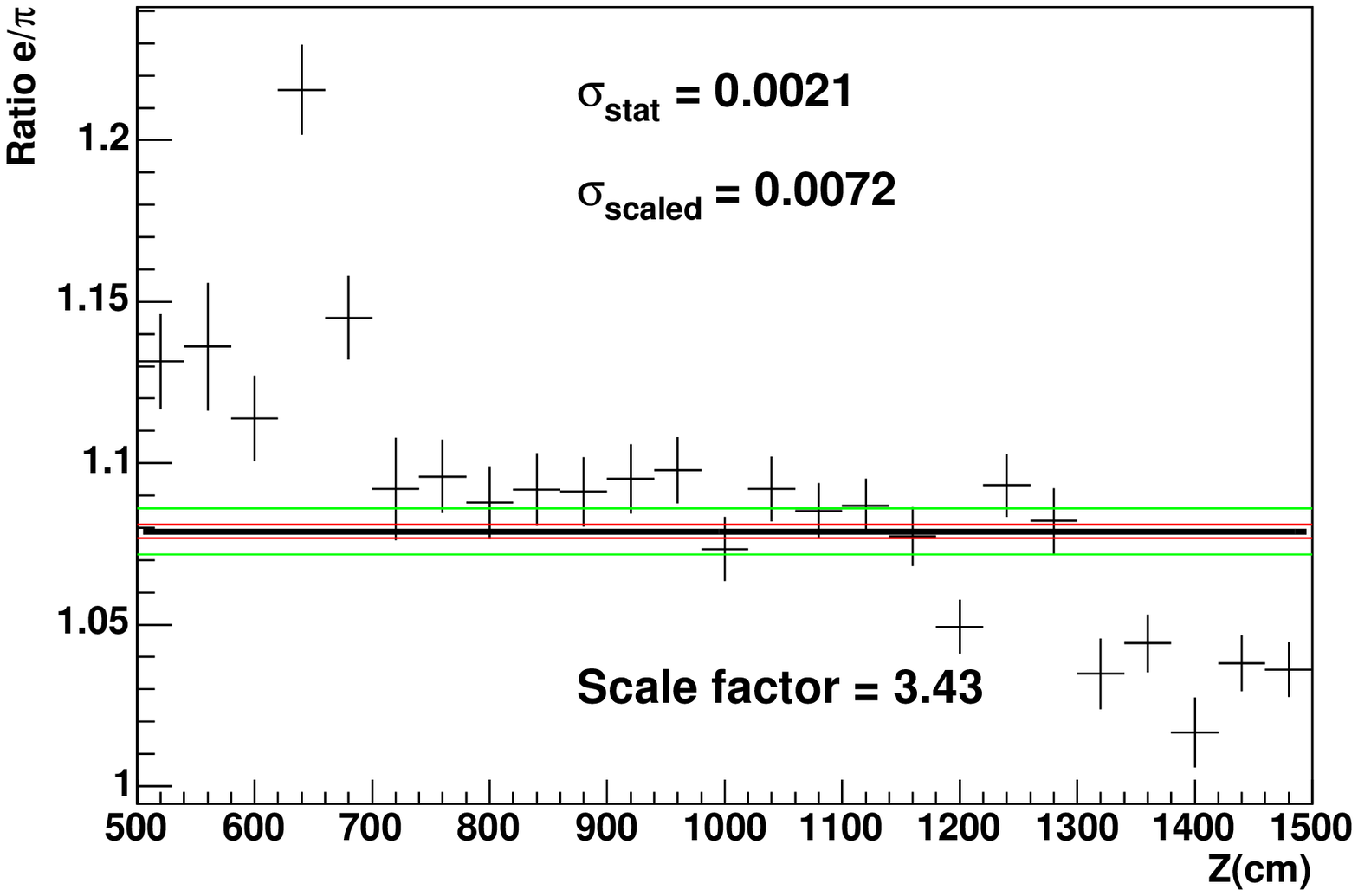} 
\begin{center}
{\small Figure 7: The  measured ratio $N_{K_{e3}}/N_{K_{\pi2}}$ versus 
$z$ coordinate  of the vertex.}
\end{center}
\end{minipage}\\
\vspace*{0.3cm}
\linebreak
The spread of the measured values around average is significantly larger than 
that expected from gaussian statistics. In extracting the systematics error we
used an approach proposed by PDG \cite{PDG06} when calculating average from
different experiments: a scale factor is defined $s=[\chi^{2}/(N-1)]^{1/2}$. 
If we scale up all the errors by this factor, the $\chi^{2}$ becomes N-1, as required
by ideal Gaussian statistics and the error of the average scales up by the same
factor. The systematic error is then defined as 
$\sigma_{syst}=\sigma_{stat}\sqrt{s^{2}-1}$. For example, the scale factor for Fig.~6
equals 3.25 and for Fig.~7 it is 3.43. Note, that Fig.~7 demonstrates clear 
systematics in the region of z before the vacuumed decay volume ($z<700$
cm) and
after it ($z>1300$ cm). We could cut out this regions, reducing systematics related to
$z$, but it does not reduce significantly the scale factors for other distributions.
That is why we decided not to introduce this ``a posteriori'' cut. 

In this way several more distributions were studied, in particular, over 
azimuthal and polar angles of the secondary track etc. The average scale factor
observed is $s \sim 3.5$. This gives estimation for systematic error in 
$Br_{K_{e3}}/Br_{K_{\pi2}}$ of 0.0014 or the systematic error in $Br_{K_{e3}}$ of $0.030 \%$. 

 A systematics related to a possible admixture of electrons, muons and pions 
in the beam was separately studied. Fig.8 shows the distribution of events from the
 $E/p$ peak of Fig.~4 over momentum $(p)$. The histogram corresponds to the selection 
of the $E/p$ peak region by a simple cut, and the points with errors are the results
of the fit of $E/p$ distribution for every bin in momentum. The absence of a bump in the 
region of the beam momentum ($\sim 25$ GeV) indicates the absence of this type of
background. Indeed, beam electrons are suppressed by two $\check{C}$-counters in the 
beam and by the $\Delta \theta> 2 $ mrad cut.

It is easy to show that the amount of electrons from the decay chain 
$\pi \rightarrow \mu \rightarrow e$ of the beam pion is negligible as compared with 
the decay chain $K \rightarrow \mu \rightarrow e$, which is correctly reproduced in our
MC: the number of pions which may pass $\check{C}$-counters ``Veto'' is at most $0.5 \%$,
i.e it is 1/6 of kaons, the lifetime factor($\gamma c \tau$) is 7.5, then taking
into account the $K \rightarrow \mu \nu$ branching of .63 we get factor of 30 in
favour of kaon decays.\\ 

\begin{center}
\begin{minipage}[h]{15.cm}
\begin{center}
\includegraphics[width=8. cm]{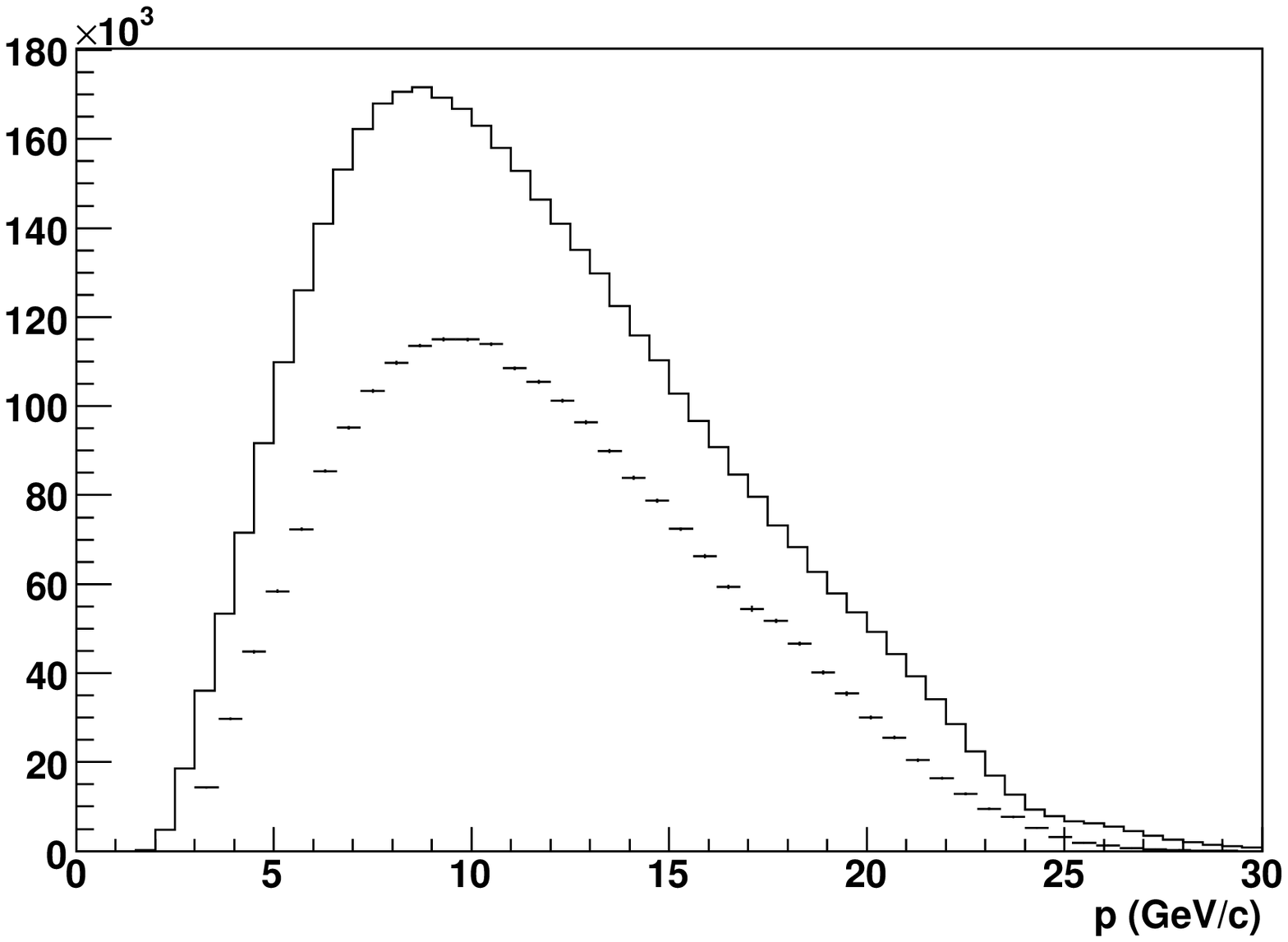}\\ 
{\small Figure 8: The  measured momentum distribution of the tracks, identified
as electrons. Histogram corresponds to the ``rough'' identification by the selection
$0.6< E/p$, the points with errors --- results of the fits of the $E/p$ distributions
for each bin in $p$.}
\end{center}
\end{minipage}\\
\end{center}
The effects on $N_{K_{e3}}/N_{K_{\pi2}}$ from the cuts variation (vertex fit probability, $\Delta \theta> 2 $ mrad, $E_{SP1} > 1$~GeV)
and different parametrization of the signal and
background of Fig.~2 -- Fig.~5
are less then estimation of systematic error from previous section.\\
Summing up all the systematics observed leads to our  final result: 
\begin{displaymath}
 Br_{K_{e3}}/Br_{K_{\pi2}}=0.2449 \pm 0.0004(stat) \pm 0.0014(syst)\ ,
\end{displaymath}
\begin{displaymath}
Br_{K_{e3}}=(5.124 \pm 0.009(stat) \pm 0.029(norm) \pm 0.030(syst))\%\ .
\end{displaymath}
The comparison of our results with the E865 \cite{Poblaguev}
 shows reasonable agreement. \\
\section{Extraction of $|V_{us}|$}
In the Standard Model the Born-level matrix element for the 
$K^{\pm} \rightarrow \pi^{0} l^{\pm} \nu$ decay modes is:
\begin{displaymath}
  M= \frac{G_{F}V_{us}}{2} 
 [f_{+}^{K^{+} \pi^{0}}(p_{K}+p_{\pi})_{\alpha}+
f_{-}^{K^{+} \pi^{0}}(p_{K}-p_{\pi})_{\alpha}]\bar u(p_{\nu}) (1+ \gamma^{5})\gamma^{\alpha} 
v(p_{l})\ , 
\end{displaymath}
here  $\frac{1}{\sqrt{2}}[f_{+}^{K^{+} \pi^{0}} \cdot (p_{K}+p_{\pi})_{\alpha} +
f_{-}^{K^{+} \pi^{0}} \cdot (p_{K}-p_{\pi})_{\alpha}]\equiv
< \pi^{0} | \bar s \gamma_{\alpha} (1- \gamma_{5})u | K^{+}> $ ; 
 $f^{K^{+}\pi^{0}}_{\pm}(t)$~--- form factors, which depend on  $t=(p_{K}-p_{\pi})^{2}=
 (p_{l}+p_{\nu})^{2}$~--- the square  of the four momentum transfer to the leptons.

The term in the vector part, proportional to $f_{-}$ is reduced(using the Dirac
equation) to an effective scalar term, proportional to $m_{l}$. That is why
in case of $\mbox{K}_{e3}$ decay one can neglect the term proportional to $f_{-}$. 

The $K^{\pm}_{e3}$ decay rate can be expressed as:
\begin{displaymath}
\Gamma(K^{\pm}_{e3})=\frac{Br_{K^{\pm}_{e3}}}{\tau(K^{\pm})}=
\frac{G_{\mu}^{2}}{384 \pi^{3}}M^{5}_{K}|V_{us}|^{2} 
|f_{+}(0)|^{2}I^{e}_{K^{+}}S_{EW}(1+\delta_{SU2}+\delta^{e}_{+})^{2}\ .
\end{displaymath}

Here $S_{EW}=1.0232 \pm 0.0003$ is the short-distance radiative correction \cite{Sirlin}; \linebreak
$\delta_{SU2}=(2.31 \pm 0.22)\%$ takes into account the
difference between $f^{K^{0}\pi^{-}}_{+}(0) \equiv f_{+}(0)$  and $f^{K^{+}\pi^{0}}_{+}(0)$
\cite{Cirigliano,Moulson};
$\delta^{e}_{+}=(0.03 \pm 0.1)\%$ is the long distance radiative correction for $K^{+}_{e3}$,
for the fully inclusive $K_{e3 (\gamma)}$ decay~\cite{Cirigliano,Moulson}. The $I^{e}_{K^{+}}$  is
the dimensionless decay phase space integral \cite{Leutwyler}:
\begin{displaymath}
I^{e}_{K^{+}}= \int_{0}^{(M_{K}-M_{\pi})^{2}} dt \frac{1}{M^{8}_K} \lambda^{3/2}
(f_{+}(t)/f_{+}(0))^{2}\ .
\end{displaymath}
Where $\lambda= (M_{K}^{2}-t-M_{\pi}^{2})^{2}-4tM_{\pi}^{2}$.

To extract $|V_{us}|$, we take $Br_{K^{\pm}_{e3}}$ from the present experiment, 
$\tau(K^{\pm})=(12.385 \pm 0.025)$~nsec~--- the charged kaon life-time from PDG06 \cite{PDG06} and calculate
$I^{e}_{K^{+}}$ 
from our measurement of $f_{+}(t)$, where the quadratic non-linearity was observed
for the first time \cite{Iou}:
\begin{displaymath}
I^{e}_{K^{+}}=0.15912 \pm 0.00084(stat) \pm 0.00114(syst)\ .
\end{displaymath}
The systematic error reflects the difference between the quadratic and linear
fit of the $f_{+}(t)$.
Putting everything together we get:
\begin{displaymath}
|V_{us}f_{+}(0)|=0.2186  \pm 0.0009_{Br} \pm 0.0012_{th}\ .
\end{displaymath}
And finally:
\begin{displaymath}
|V_{us}|=0.2275 \pm 0.0009_{Br} \pm 0.0022_{th}\ .
\end{displaymath}
If theoretical value $f_{+}(0)=0.961 \pm 0.008$ \cite{Leutwyler}  
is used.
\section{Summary and conclusions}
The $K^{-}_{e3}$ decay has been studied using in-flight decays of 25 GeV 
$K^{-}$, detected by ISTRA+ magnetic spectrometer. Due to the high
statistics, adequate resolution of the detector and good sensitivity over
all the Dalitz plot space, the  errors are significantly reduced
as compared with the previous measurements. 
 The $K_{e3}$ branching is measured to be: 
\begin{center} 
 $Br_{K_{e3}}/Br_{K_{\pi2}}= 0.2449 \pm 0.0004(stat) \pm 0.0014(syst) $.
\end{center}
\begin{center} 
$Br_{K_{e3}}=(5.124 \pm 0.009(stat) \pm 0.029(norm) \pm 0.030(syst) )\%$.
\end{center}
From that  we obtain:
\begin{center} 
 $|V_{us}f_{+}(0)|= 0.2186 \pm 0.0009_{Br} \pm 0.0012_{th} $.
\end{center}
    Which leads to: 
\begin{center} 
$|V_{us}|=0.2275 \pm 0.0009_{Br} \pm 0.0022_{th}$
\end{center}
if the theoretical value for $f_{+}(0)$ is substituted.
Our result on $|V_{us}|$ is in reasonable agreement with that from 
charged \cite{Poblaguev} and neutral \cite{KTeV},\cite{KLOE1},\cite{KLOE2} kaon decays.\\

The work is  partially supported by the  RFBR grants N07-02-00957, N07-02-16065 and by the  grant of the Russian Science Support
Foundation. \\

\end{document}